\documentclass[epj]{svjour}

\usepackage{epsfig,psfrag}
\usepackage{dcolumn}     
\usepackage{bm}          
\usepackage{amssymb}          
\usepackage{times}
\usepackage{graphics}

\newcommand{\beq}{\begin{equation}}
\newcommand{\eeq}{\end{equation}}
\newcommand{\beqn}{\begin{eqnarray}}
\newcommand{\eeqn}{\end{eqnarray}}
\newcommand{\bra}{\left<}

\newcommand{\ket}{\right>}

\newcommand{\bfig}{\begin{figure}}
\newcommand{\efig}{\end{figure}}
\newcommand{\bmp}{\begin{minipage}}
\newcommand{\emp}{\end{minipage}}
\newcommand{\bc}{\begin{center}}
\newcommand{\ec}{\end{center}}

\begin{document}
\title{Phonon-induced decoherence and dissipation in donor-based charge qubits}
\author{J.\ Eckel, S.\ Weiss, and M.\ Thorwart}
\institute{Institut f\"ur Theoretische Physik IV,
Heinrich-Heine-Universit\"at D\"usseldorf, D-40225 D\"usseldorf,
Germany}
\mail{Jens Eckel, \\ \email{eckelj@thphy.uni-duesseldorf.de}}

\date{\today}

\abstract{
We investigate the phonon-induced decoherence and dissipation in a
donor-based charge quantum bit realized by the orbital states of an 
electron shared by two dopant ions which are implanted in a 
silicon host crystal. The dopant ions are taken from the group-V
elements  Bi, As, P, Sb. The excess electron is 
coupled to deformation potential acoustic phonons which dominate in
the Si host. The particular geometry tailors a non-monotonous frequency
distribution of the phonon modes. We determine the exact qubit dynamics 
 under the influence of the phonons by employing 
the numerically exact
quasi-adiabatic propagator path integral scheme thereby taking
into account all bath-induced correlations. In particular, we
have improved the scheme by completely eliminating the Trotter
discretization error by a Hirsch-Fye extrapolation. By comparing 
the exact results to those of a Born-Markov 
approximation we find that the latter yields appropriate estimates for
the decoherence and relaxation rates. However, noticeable quantitative
corrections due to non-Markovian contributions appear.}
\maketitle
\section{Introduction}
During the last decade it turned out that solid-state based nano structures 
are promising candidates 
for the realization of quantum information processing devices \cite{njpFocus}.
The building blocks are quantum mechanical two-state systems (qubits) and 
some of the proposed designs have been realized 
in groundbreaking experiments, see Ref.\ \cite{njpFocus} for a recent
review on this field. Thereby, various approaches have been undertaken, ranging
from superconducting flux and charge qubit devices to devices using the spin or
the charge degrees of freedom of individual electrons in confined geometries.
Aiming at an extreme miniaturization of solid-state devices down to the
nm-scale, it has been proposed to implant individual dopant atoms in a
semiconductor crystal and to use nuclear spin states of buried phosphorus
dopants to realize a spin-qubit (Kane's proposal \cite{kane98}). Complementary
to the Kane architecture, the charge degree of freedom of a
single electron shared by two donor atoms in a host crystal can be used  
for the coding of the logical information, as proposed in Ref.\  
\cite{barret03,hollenberg04}. 
Thereby, the logical
states $|0\rangle$ and $|1\rangle$ are realized by the charge
states of the double-donor-system  with the excess electron either located on
the left or on the
right donor. The transition between these states occurs via tunneling of the electron
between the two dopants. The charge qubits can in principle be controlled
efficiently by external electric fields, e.g., by an applied gate voltage.
This property renders the proposed architecture attractive for realizing control
schemes with available fabrication and read-out
technologies \cite{hollenberg04}. Experimental progress for this kind
of ion-implanted Si:P nanostructures has been reported recently 
\cite{McCamey}. 

On the other hand, solid-state qubits suffer from the large number of degrees
of freedom due to their embedding in a complex many-particle environment. 
The  environmental  decoherence and dissipation lead to a deterioration of the
performance of quantum logic operations and also strongly influence entanglement
between qubits \cite{thorwart02} necessary for quantum gate operations. 
 Various sources of decoherence include nuclear spins, phonons, and
electromagnetic fluctuations in the host crystal. To gain a detailed
understanding of the various decoherence
mechanisms, realistic model calculations have to be performed which then
allow to sort out the different contributions. In this work we concentrate 
on the influence of a phonon bath on the shared electron. To be definite, we
consider a charge qubit formed by two group-V donors as proposed in Ref.\ \cite{koiller06}. 
One donor is formed by a
phosphorus atom while the second donor  will be one of the class $\{$Bi,
As, P, Sb$\}$. The donor pair is assumed to be implanted in a silicon crystal
host and share a common electron. We consider linear acoustic phonons coupled to
the electron and determine the dynamics of the charge oscillations between the
two donors. Due to the particular geometry, a tailored phonon environment is
formed for the electron which depends non-monotonously on the phonon
frequencies. In order to provide accurate quantitative results on the
decoherence and dissipation rates, we apply the numerically exact iterative quasi-adiabatic
propagator path integral (QUAPI) scheme \cite{QUAPI,Tho98}. In particular, we
have improved the widely used method by providing a recipe to completely
eliminate the Trotter discretization error. This allows to obtain fully
convergent exact results by extrapolation to a vanishing Trotter
increment  
\cite{fye86}.  An appealing alternative to extensive numerical studies are
approximative calculations which, for instance, rely on the weak coupling
between the qubit and the environment. The most familiar Born-Markov or
weak-coupling approximation (WCA) \cite{weiss} yields to simple closed
expression for the decoherence and relaxation rates. However, they apply for
typical situations when the bath has a smooth frequency distribution 
\cite{weiss}. In our case, the environment is particularly shaped by the
geometry leading to a non-monotonous bath spectral density. Hence, it is not 
{\em a priori\/} clear whether the widely used WCA is appropriate and a careful
check is desirable. By comparing the exact numerical QUAPI results with the
approximate WCA results below, we will show that for realistic parameters, the
WCA typically yields the correct order of magnitude for the decoherence and
relaxation rates. However, differences are noticeable when a
quantitative comparison is made. We furthermore note that the calculated phonon
decoherence and relaxation rates comprise a fundamental upper limit for the
coherence properties of this architecture which can hardly be overcome. 

The presented set-up is related to a double-quantum dot charge qubit realized in
a GaAs semiconductor \cite{thorwart05}, where the
geometrical constraints induce charge qubit oscillations with noticeable
non-Markovian corrections due to the particularly shaped phonon environment.
While  piezoelectric phonons dominate in GaAs, we have to consider here the dominating deformation
potential electron-phonon coupling since the Si crystal displays inversion
symmetry. 
\section{The model}
To study the influence of the phonons on decoherence and dissipation, we 
assume that the charge qubit is isolated from any leads. 
It is formed by a pair of donor atoms embedded in a silicon substrate, which share a 
single excess electron 
\cite{barret03,hollenberg04}. To be specific, we consider the situation of one donor being a 
phosphorous atom while the second one is an individual donor atom $X$ chosen 
from the group $X\in\{$Bi, As, P, Sb$\}$ \cite{koiller06}. Then, the two logic states $|0\rangle,|1\rangle$ 
of the charge qubit are defined by the electron residing either at donor $1$ 
or $2$, respectively. 
The total Hamiltonian is given in terms of the standard spin-boson
model 
\cite{weiss,leggett_rmp}
\beq
H=H_S+H_B+H_{SB}\,,
\label{eq:ham}
\eeq
where $H_S$ is the two-state Hamiltonian for the charge qubit, $H_B$ models the 
phonon bath and $H_{SB}$ includes the electron-phonon coupling.

\subsection{Model for the charge qubit}
We represent the Hamiltonian of the charge qubit in the basis of the two localized electronic states 
 denoted as $|L\rangle\equiv|0\rangle$ and $|R\rangle\equiv|1\rangle$, each being the $1s$ orbital
 of the left/right donor atom, the latter being placed at the origin and at the position 
$d{\mathbf e}_y$, see Fig.\ \ref{fig:geometry} for a sketch of the geometry. 
 The localized orbital belonging 
to the right (left) donor is fully described by the position vector of the
electron, i.e., 
 ${\mathbf r}_L={\mathbf r}$ and ${\mathbf r}_R={\mathbf r}+d{\mathbf e}_y$, respectively.  
In addition, we allow for an external constant energy bias $\epsilon$ which for instance
could be due to a nearby capacitive gate. In terms of the Pauli spin matrices
$\sigma_i$, the two-state Hamiltonian then reads 
\beq
H_S = \hbar \Delta \sigma_x +\hbar \epsilon\sigma_z\,.
\label{eq:Hs}
\eeq
The two lowest lying energy eigenstates $|E^{\pm}\rangle$ are given as an (anti-)symmetric superposition of 
the localized states $|L\rangle$ and $|R\rangle$ such that
$|E^{\pm}\rangle=(|L\rangle\pm|R\rangle)/\sqrt{2}$ with energies $E_{\pm}=\mp \Delta/2$. 
The tunneling amplitude then follows as $\Delta=E_{+}-E_-$ and is a function of the 
donor distance $d$.

In order to determine the tunneling amplitude $\Delta$, 
we have to calculate approximate eigenvalues of the lowest
symmetric and antisymmetric energy-eigenstate. In principle, rather highly elaborated 
methods are available for their calculation, including the anisotropic conduction band dispersion of silicon, 
the valley orbit interaction and valley interference effects \cite{koiller06,wellard05}. The latter
leads to an oscillatory behavior of the tunneling amplitude $\Delta$ for 
increasing the donor distance $d$. Noticeably, the oscillations are weak 
if the two donors are placed in the $[100]$-plane of the Si host
\cite{koiller06}. However, we aim for a detailed
and quantitative understanding of the electron-phonon decoherence mechanism and
thus resort to the simplest  
straightforward procedure to determine the tunneling amplitude which is the
well established linear combination of atomic 
orbitals (LCAO) \cite{barret03,slater}. This tight-binding method  is very successful 
for determining the molecular orbitals 
for the $H_2^+$-molecule but can easily be
generalized to our model by introducing an effective Bohr radius \cite{ning71}. 
When we neglect the conduction-band anisotropy, we can assume that the localized states
$\left|\xi\ket$ ($\xi=L,R$) are represented by the $1s$ orbital of each donor. They read
\beq
\left|\xi\ket=\sqrt{\frac{1}{\pi a_{\xi}^3}}e^{-r_\xi/a_{\xi}}
\label{locstat}
\eeq
where $a_{\xi}$ is the effective Bohr-radius of the donor $\xi$ \cite{ning71} and 
$r_\xi=|{\mathbf r}_\xi|$.
In the following, the left donor is assumed to be the phosphorous atom, 
whereas the right donor is taken from the group-V donors $\{$Bi, As, P, Sb$\}$. 
Hence, we introduce the ratio $p$ such that $a_R=pa_L$. 

To calculate the energy levels an ansatz for the wave function for the (anti-)symmetric ($\mp$)
part is made and the overlap between the two wave functions is calculated, yielding the energies for
the (anti-)symmetric state. If energies are scaled in atomic units, they read \cite{slater}
\beq
E_{\pm}=E_1^{(\pm)}(d)+E_2^{(\pm)}(d)\,.
\eeq
Here, $E_1^{(\pm)}(d)$ is the kinetic energy and $E_2^{(\pm)}(d)$ is the potential energy, both being functions of the (dimensionless) donor distance $d$. They read 
\beqn
E_1^{(\pm)}(d)&=&\frac{1\pm e^{-d}(1+d-d^2/3)}{1\pm e^{-d}(1+d+d^2/3)}\nonumber\\
E_2^{(\pm)}(d)&=&-2\frac{1\pm 2\,e^{-d}(1+d)+(1/d)-(1/d+1)\,e^{-2d}}{1\pm e^{-d}(1+d+d^2/3)}\,. 
\nonumber \\
\eeqn
Due to the fixed positions of the donors, there is no need to minimize the energy with respect to the donor distance, in contrast to 
analogous calculations for the $H_{2}^+$-molecule. According to the LCAO
calculations typical tunneling amplitudes for a distance of $d=7.06$ nm (which
corresponds to a separation of the two dopants by $n=13$ lattice sites) follow
as $\Delta\approx 16$ meV. This is qualitatively consistent with the results
obtained from a more refined approach taking into account interference effects
in the Si band structure\cite{koiller06}.  

\subsection{Coupling to linear acoustic phonons}

The phonon bath is due to the silicon host crystal and is modeled as usual in terms of the bosonic operators $b_{\mathbf q}$ as 
\beq
H_B=\hbar \sum_{\mathbf q} \omega_{\mathbf q} b^\dagger_{\mathbf q} 
b_{\mathbf q} \, , 
\eeq
with the phonon dispersion relation $\omega_{\mathbf q}$.
The electron-phonon interaction reads \cite{mahan,brandes99}
\begin{equation}
H_{SB} = \hbar \sum_{\mathbf q} (\alpha_{\mathbf q}^{L} N_L + 
\alpha_{\mathbf q}^{R} N_R ) (b^\dagger_{\mathbf q}+b_{-\mathbf q}) \,.
\end{equation}
Here, $N_\xi=0,1$ is the number of the excess electrons on the
donor $\xi$, respectively,  and $\alpha_{\mathbf q}^{\xi} = \lambda_{\mathbf q}
e^{-i {\mathbf q} \cdot {\mathbf r}_{\xi}} F_\xi ({\mathbf q})$. 
The coupling constant $\lambda_{\mathbf q}$ depending on the 
wave vector {\bf q} is specified below. 
Note that the phonons can propagate in all three dimensions, and
the electron-phonon coupling is not isotropic in general \cite{Eduardo04}.
To take care of the charge distribution in each donor we define a form factor 
according to
\beq
F_\xi ({\mathbf q}) = \int d^3r \, n_\xi({\mathbf r}) e^{-i{\mathbf q}
\cdot {\mathbf r}}\,,
\label{eq:formfactor}
\eeq
where $n_\xi({\mathbf r})$ is the charge density  of the donor $\xi$.
The coupling Hamiltonian is rewritten in the form \cite{brandes99}
\beq
H_{SB} = \frac{\hbar}{2} \sigma_z \sum_{\mathbf q} g_{\mathbf q} 
(b^\dagger_{\mathbf q}+b_{-\mathbf q}) \, , 
\eeq
with $g_{\mathbf q}=\left[\lambda_{\mathbf q}(F_L({\mathbf q})-F_R({\mathbf q}))\right]$.
The charge density distribution then follows directly from Eq.\ (\ref{locstat}) as 
$n_{\xi}({\mathbf r})=\left|\bra{\mathbf r}|\xi\ket\right|^2$, 
which leads to the form factors  
$F_{L}({\mathbf q}) = f_L({\mathbf q})$ and 
$F_{R}({\mathbf q}) = f_R({\mathbf q})e^{-i{\mathbf q}\cdot d{\mathbf e}_y}$ with 
$f_\xi({\mathbf q})=16/[4+(qa_\xi)^2]^2$. 

In this work we focus on linear acoustic phonons with linear dispersion relation 
$\omega_{\mathbf q} = s |{\mathbf q}|$, $s$ being the sound velocity 
for silicon ($s\approx 9\times 10^3\,$m/s) \cite{nsm}. Since the silicon crystal has an inversion center there is no piezoelectric coupling between 
electrons and phonons, wherefore the dominating coupling is due to the deformation potential. 
Thus, the coupling constant reads
\beq
\lambda_{{\mathbf q}} = \frac{D}{\hbar}q\sqrt{\frac{\hbar}{2\rho_mV\omega_{{\mathbf q}}}}\,,
\label{eq:coupl_const}
\eeq
where $D$ is the deformation constant for silicon ($D\approx 8.6\,$eV, see Ref.\ \cite{friedel}), $\rho_m$ is the mass density of silicon 
($\rho_m\approx 2.33\times 10^3\,$kg\ m$^{-3}$, see Ref.\  \cite{nsm}) 
and $V$ is the volume of the unit cell.

All the properties of the phonon bath can be captured in 
the spectral density defined as
\beq
G(\omega) = \sum_{\mathbf q} |g_{\mathbf q}|^2 \delta(\omega - \omega_{\mathbf q}) \, . 
\eeq
Using Eq.\ (\ref{eq:coupl_const})  and the definition of the form factors and   
taking into account the geometry, the sum over ${\mathbf q}$ can be transformed into a continuous integral which can readily be carried out. One then obtains the spectral density
\beqn
G(\omega) &=& \frac{64D^2}{\pi^2\rho_m\hbar s^5}\omega^3\\
&&\times\left[\left(4+\left(\frac{\omega}{s}a_L\right)^2\right)^{-4}+\left(4+\left(\frac{\omega}{s}pa_L\right)^2\right)^{-4}\right.\nonumber\\
&&-\left.2\left(4+\left(\frac{\omega}{s}a_L\right)^2\right)^{-2}\left(4+\left(\frac{\omega}{s}pa_L\right)^2\right)^{-2}\right.\nonumber\\
& & \left. \times j_0\left(\frac{\omega}{s}d\right)\right]\,.\nonumber
\label{eq:specdens_si}
\eeqn 
where $j_0$ is the spherical Bessel function. The spectral density is sketched in the inset of Fig.\ \ref{fig:specdens_si}.
The low-frequency behavior is superohmic according to
 $G(\omega\rightarrow 0)\propto\omega^3$,
while in the high-frequency limit, it decays algebraically as 
$G(\omega\rightarrow\infty)\propto\omega^{-5}$. 
The crossover between these two limits occurs on a frequency scale $\omega_c = s/a_P\equiv\tau_c^{-1}$,
where $a_P$ is the radius of the phosphorus donor ($a_P=1.22\,$nm, see Ref.\  \cite{ning71}), 
yielding $\omega_c=2.46\,$THz, which corresponds to an energy of $10.17\,$meV. 
As we will see below, typical tunneling
amplitudes $\Delta$ are comparable to this energy scale. Thus, the 
frequency distribution of the bath is no longer monotonous in the range of the relevant system energies. 
As common approximative analytical treatments \cite{barret03,weiss} of phonon-induced decoherence 
typically involve a smooth frequency distribution, it is not {\em a priori} clear whether their results are applicable to this situation. Moreover, the used Born-Markovian 
approximation which neglects bath-induced correlations might not describe properly the 
dynamics. This can be seen from the autocorrelation function \cite{weiss} of the bath, i.e., 
\beqn
L(t) & = & L_R(t)+iL_I(t) \nonumber\\[0.2cm]
& = & \frac{1}{\pi} \int\limits_0^\infty d\omega G(\omega) \left[ \coth
\frac{\hbar \omega \beta}{2} \cos \omega t - i \sin \omega t
\right], \label{eq:response}
\eeqn
which  is shown in Fig.\ \ref{fig:specdens_si}. The typical width of the correlation function is comparable to the time scale $\Delta^{-1}\approx\omega_c^{-1}$ of the system dynamics. 
The Born-Markov approximation corresponds to replacing the strongly
peaked real part $L_R(t)$ by a $\delta$-function with the
corresponding weight while the imaginary part $L_I(t)$ is often
neglected. However, since the geometry tailors a specific structured phonon environment 
for the charge qubit, it is not clear from the very beginning that the Markovian assumption is valid. 
It is the main purpose of this work to investigate this issue and compare exact real-time path integral simulations with approximate weak-coupling (Born-Markov) results.

\section{The improved QUAPI scheme}

The dynamics of the charge qubit is described in terms of the time
evolution of the reduced density matrix $\rho(t)$ which is obtained 
after tracing out the bath degrees of freedom, hence
\beqn
\rho(t) & = & {\rm tr}_B K(t,0) W(0) K^{-1} (t,0)\, , \nonumber \\
K(t,0) & = & {\cal T} \exp \left\{ -\frac{i}{\hbar} \int\limits_0^t 
dt' H \right\} \, . 
\label{eq:rhored}
\eeqn
$K(t,0)$ denotes the propagator of the full system plus bath, 
$\cal T$ is the time ordering operator and $H$ is the Hamiltonian, Eq.\ (\ref{eq:ham}). 
The full density operator $W(0)$ at initial time $t=0$ is as usual assumed  
factorizing according to  
$W(0) \propto \rho(0)
\exp [-H_B/(k_B T)]$, where $k_B$ is the Boltzmann constant and $T$ is the 
bath temperature \cite{weiss}. In this work the qubit
dynamics always evolves from the initial state 
$\rho(0) =\left|L\ket\bra L\right|$.  

In order to investigate the dynamics of the system, we use the 
quasi-adiabatic propagator path integral (QUAPI) scheme 
\cite{QUAPI} being a numerically exact iteration scheme which has been successfully 
adopted to many problems of open quantum systems \cite{Tho98,thorwart05}.
For details of the iterative scheme we refer to  previous works \cite{QUAPI,Tho98} and do 
not reiterate the scheme here again. 
However, we have improved the method at one important step and we will 
describe this in greater detail next.  

The algorithm is based on a symmetric Trotter splitting of the short-time
propagator $K(t_{k+1},t_k)$ of the full system into a part depending on $H_S$ and $H_B+H_{SB}$
describing the time evolution  on a time slice $\Delta t$. This is exact in the limit $\Delta t \to 0$
but introduces a finite 
Trotter error to the propagation which has to be eliminated by choosing $\Delta t$ small 
enough that 
convergence has been achieved. On the other side, the bath-induced correlations being 
non-local in time 
are included in the numerical scheme over a finite memory time $\tau_{mem}=K\Delta t$ 
which roughly corresponds to the time range over which 
the bath autocorrelation function $L(t)$ given in Eq.\  (\ref{eq:response})  
is significantly different from zero. Note that for any finite temperature $L(t)$ 
decays exponentially at long times \cite{weiss} justifying this approach. 
To obtain convergence with respect to the memory time, $K$ has to be increased until 
converged results have been found. 
However, the numerical effort grows exponentially with the memory length $K$ and for 
the present two-level system, the memory length is restricted to typical values of 
$K=12$ on a standard processor with 
2 GB RAM for practical reasons.

Thus, the two strategies to achieve convergence are countercurrent. To solve this, 
the principal of least dependence 
has been invoked \cite{Tho98} to find an optimal time increment in between the two 
limits. However, here we show that the algorithm can be improved by applying a 
different strategy. 

We first choose some small enough time increment $\Delta t$. Then, one has to 
increase the memory time $\tau_{mem}$ by increasing $K$ until convergence has 
been achieved. Typical results of this memory convergence check are shown in 
Fig.\  \ref{fig:cut-off}. Shown is the decoherence rate $\gamma$ for increasing 
memory time for different donor distances ($p=1$) for the symmetric qubit 
$\epsilon=0$. Note that the decay rate has been obtained by fitting the 
results for the population difference $P(t)=\langle \sigma_z\rangle_t$ to an 
exponentially decaying cosine. The remaining error is the Trotter error. 
However, following Ref.\  \cite{fye86}, 
for any Hermitian observable, this symmetric Trotter error vanishes 
quadratically in the limit $\Delta t \to 0$. 
This opens the possibility to extrapolate the results to $\Delta t \to 0$, 
thereby completely eliminating the Trotter error. This is done by 
decreasing $\Delta t$ from the initial value and then by finding the extrapolated 
exact result (of course, convergence has to be verified again for the 
smaller values of $\Delta t$). Typical results of this extrapolation 
procedure are shown in Fig.\ \ref{fig:trotter}, indicating that the 
numerical values follow
a line for decreasing step sizes. Note that we consider $P(t_{fix})$ 
at an arbitrary time $t_{fix}=34.1 \omega_c$ in this example. 
Indeed, we find the predicted behavior for the Trotter error to vanish 
and perform a linear 
regression to $\Delta t\to0$, also shown in Fig.\ \ref{fig:trotter}. 
The $y$-axis intersection gives the numerical exact value for the observable 
of interest, in this case 
afflicted with a tiny error bar coming from the linear regression. 
In general, the convergence properties of an observable strongly depend on 
the involved parameters, similar to path-integral quantum Monte-Carlo 
simulations \cite{Weiss}. Different 
observables show different behaviors with decreasing Trotter step size 
$\Delta t$, as for instance the particle density in contrast to the 
energy of the system in Ref.\ \cite{Weiss}. 

\section{The dynamics of the charge qubit}

Equipped with the numerically exact improved QUAPI scheme, we can now study the dynamics of the charge qubit in detail. 
To extract the 
decoherence rate $\gamma$, the relaxation rate $\gamma_r$, the equilibrium population difference 
$P_{\infty}$ and the oscillation frequency $\Omega$, we fit
a combination of exponentially decaying cosine  and sine functions  
\cite{weiss} to the numerically exact data, from which the Trotter error has been
eliminated. 
We can then investigate the dependence of the above quantities 
on the experimentally relevant parameters. 
We emphasize again that realistic assumptions on the geometry of the system enter the spectral density
Eq.\ (\ref{eq:specdens_si}) and thus allow to calculate quantitative realistic results.

One of the major goals of this work is to verify the Born-Markov (weak-coupling)
approximation, 
since the later results in very simple 
and compact formulas for parameters governing the dynamics. 
Hence, we compare the exact QUAPI results  with
results obtained within a WCA which are known as \cite{weiss}
\beqn  
\gamma&=&\frac{\gamma_r}{2}+\frac{2\pi\epsilon^2}{\Delta_b^2}\,S(0)\,, \label{eq:gam} \\
\gamma_r&=&\frac{\pi\Delta_{\rm eff}^2}{2\Delta_b^2}\,S(\Delta_b)\,
,\label{eq:gamma_r}\\
\Omega^2&=&4\Delta_{\rm eff}^2[1-2 \, {\rm Re } \, u(2
i\Delta_b)]+4\epsilon^2\,,\label{eq:freq}\\
P_{\infty}&=&-\frac{\epsilon}{\Delta_b}\tanh\left(\frac{\hbar\Delta_b\beta}{2}\right)\,.\label{eq:equil_wca} 
\eeqn
The spectral function $S(\omega)$, related to the phonon spectral density, Eq.\ (\ref{eq:specdens_si}), via 
$S(\omega)=G(\omega)\coth(\hbar\omega/(2k_B T))$, represents emission and absorption of a single phonon and 
$\Delta_b=2\sqrt{\Delta_{\rm eff}^2+\epsilon^2}$ is twice the effective
qubit frequency.
$\Delta_{\rm eff}$ is the effective tunnel matrix element at $T=0$
\cite{weiss}, which includes the renormalization by a Franck-Condon factor
stemming from the high-frequency  modes of the reservoir \cite{weiss}. 
For our case, one easily finds that $\Delta_{\rm eff}\approx \Delta$ with a
deviation of less than $1 \%$. The function $u(z)$ is 
defined in terms of the frequency integral
\beq
u(z)=\frac{1}{2}\int\limits_0^{\infty}d\omega\,\frac{G(\omega)}{\omega^2+z^2}\left(\coth\left(\frac{\hbar\omega}{2 k_B T}\right)-1\right)\,.
\eeq

\subsection{Coherent charge oscillations for the symmetric qubit $\epsilon=0$}

For the symmetric qubit with zero bias (i.e., only decoherence, no dissipation), we have calculated the time evolution of $P(t)$ and have observed coherent charge oscillations. In order to quantify them, we define the quality factor 
$Q=\Omega/(\pi\gamma)$ where the frequency $\Omega$ and the decoherence rate $\gamma$ have been been obtained from the fit as described above. We have performed extensive simulations for three different donor distances $d$ for various combinations of donor atom species and show the results as a function of the tunneling amplitude $\Delta$ in Figs.\ 
\ref{fig:qfac1}, \ref{fig:qfac2}, and \ref{fig:qfac3}, each for a fixed
donor distance $d$. 
A variation of $\Delta$ for a fixed donor distance can, for instance, be
achieved by a small additional gate voltage which slightly distorts the $1s$
orbitals leading to an increased overlap of the wave functions.

For the smallest donor distance $d=4.34$ nm, we observe in Fig.\ \ref{fig:qfac1}
that $Q$ increases monotonously for increasing $\Delta$. 
Thereby, the results for $Q$ vary over two orders of magnitude for the different donor species at large $\Delta$. Moreover, the combination of two P donors or of one P and one (very similar) Sb donor displays the best decoherence properties. The dashed lines in Fig.\ \ref{fig:qfac1} display the results of the WCA given in Eqs.\  (\ref{eq:gam}) and (\ref{eq:freq}). A reasonable agreement is found in this case.

For intermediate donor distance $d=7.06$ nm, see Fig.\ \ref{fig:qfac2}, $Q$ first decreases but then increases again 
with increasing $\Delta$. This can be understood by the fact that $d$ determines the shape of the spectral density and,  in particular, the location of the frequency of the cross-over, relative to the qubit frequency $\Delta$. For the overall performance, the similar observation as for the smaller distance (see above) apply. Also in this case, the WCA seems to be appropriate although small 
deviations for all $\Delta$ can be observed which can be attributed to small non-Markovian corrections stemming from the specifically tailored phonon environment. 

In the case of large donor distance $d=10.32$ nm, see Fig.\ \ref{fig:qfac3} 
the differences between the various donor species almost vanish and are only noticeable 
at small $\Delta$. Also the WCA agrees well at large $\Delta$ and also yields the correct 
order of magnitude for small $\Delta$ although differences become noticeable in this regime. 
Note that in this case, $Q$ decreases for increasing $\Delta$, in contrast 
to the case of small and intermediate distances. 
 
Noticeably, we find that the $Q$-factor is independent of temperature for all relevant parameter 
combinations (not shown here). This is due to the fact that realistic temperatures correspond to 
frequencies of $T=6.5\times10^9$ Hz and hence all system frequencies are much larger. 
This behavior is in contrast to what we have recently 
reported in GaAs DQD systems \cite{thorwart05}.

Note that 
the oscillatory behavior of $\Delta$ for increasing $d$ \cite{koiller06} is not
included in this
simple LCAO approximation. However, when considering the $Q-$factor in Figs.
\ref{fig:qfac1}, \ref{fig:qfac2}, and \ref{fig:qfac3}, the oscillatory
behavior of $\Delta$ for growing donor distances
$d$ does not affect $Q$ substantially. This can be rationalized by
considering the weak-coupling results Eq. (\ref{eq:gam}-\ref{eq:freq}) for
$\epsilon=0$. Then, it becomes clear that the only part where $\Delta(d)$
appears is
in the high-frequency part of $G(\omega)$ (assuming low temperature such
the the $\coth$ approaches one and being interested in $\Delta \approx
\omega_c$). The prefactors, which in principle contain $\Delta(d)$, drop
out when the ratio is calculated.   

\subsection{Dynamics of the biased charge qubit $\epsilon \neq 0$}
When a finite bias $\epsilon \neq 0$ is present, in addition to decoherence also relaxation occurs to a non-zero asymptotic value $P_\infty\ne 0$. The corresponding decoherence and relaxation rates are also influenced by the presence of a bias in the sense that the effective qubit frequency $\Delta_b$ grows with increasing $\epsilon$. Then, the behavior of the environmental frequency distribution is essential: if it grows with increasing frequency, decoherence and dissipation will become more effective and if it decreases the environmental effects will diminish. This is what we observe from the results shown in Fig.\ \ref{fig:bias}. For comparison, we also show the corresponding WCA results, which yield the qualitatively correct behavior while differences in the quantitative results occur.

\section{Conclusion}
To summarize, we have investigated the phonon-induced decoherence and dissipation in donor-based charge qubits formed by a pair of donor atoms placed in a Si crystal host. The donor pair is formed by one P donor and one donor of the group  
Bi, As, P, Sb. We have employed the numerically exact quasi-adiabatic path-integral propagator in its iterative version. The major achievements of our work is twofold: (i) We have first improved the QUAPI scheme in the sense that the Trotter discretization error can now be completely eliminated by extrapolating the results to vanishing Trotter increment, as it is known that the error vanishes quadratically. (ii) Beyond these methodical aspects, we have obtained numerically exact results for the real-time dynamics of charge qubits under the influence of acoustic deformation potential phonons. Realistic assumptions on the
tunneling amplitude enter via LCAO calculations of the 
wave functions and the qubit energies in our model. 
Moreover, we have included the particular phonon environment tailored by 
the particular geometry of the set-up via geometrical form factors and
materials characteristics. No fitting parameters of any sort were
utilized.
 
In the absence of a static bias we have investigated the $Q$-factor of the charge oscillations 
as a function
of the donor distance and as well as a function of the tunneling amplitude. 
We have compared our results with those
obtained from a WCA within an analytical
approach in terms of real-time path-integrals and found that only small 
non-Markovian corrections appear. This can be attributed to the 
dominating super-Ohmic properties of the phonon environment at small frequencies. 
Furthermore we have investigated the dynamics in the case of a static bias and have found that the qualitative behavior of the decoherence and damping rates follows the form of the environmental frequency distribution. Non-Markovian corrections are also found in this case. 

At present, no experimental realizations of this setup is yet reported. Nevertheless, we emphasize 
that our results on the decoherence and dissipation induced by the electron-phonon
coupling represent a fundamental 
upper limit to the coherence of such donor based charge qubits which can hardly be 
negotiated due to its intrinsic nature. This has to be seen in view
of the DiVincenzo criteria \cite{DiVincenzo00} and also
for the future realization of quantum information processes. However, the dominating source of decoherence in this kind of qubit realization has to be investigated in realistic devices. 

\section*{Acknowledgments}
This work was supported by the ESF network INSTANS, the DFG-SFB Transregio 12, and the DFG Priority Program 1243.


\bfig
\bc
\epsfig{file=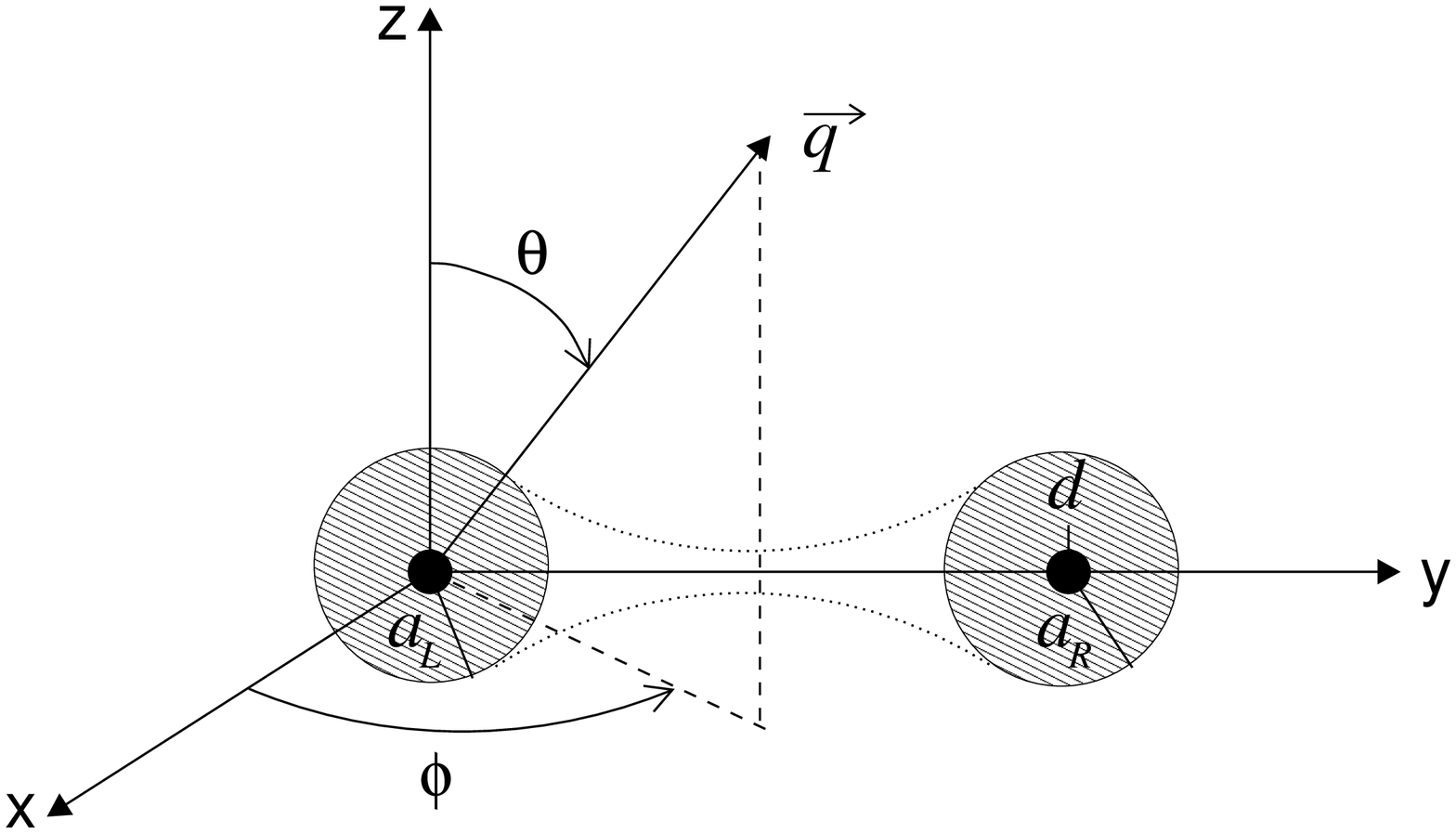, width=8.5cm, keepaspectratio=true}
\caption{Sketch of the geometry of a donor-based charge 
qubit formed two donor atoms at a distance $d$ and the various angles of the phonon propagation.
\label{fig:geometry}}
\ec
\efig

\bfig
\bc
\epsfig{file=fig2.eps, width=8.5cm, keepaspectratio=true}
\caption{The bath autocorrelation function (response function)
$L(t)=L_R(t) + i L_I(t)$ for the spectral density $G(\omega)$ (inset)
of the phonon bath for the case of two P donors in a Si host  ($p=1$, deformation 
potential phonons), 
with $s=9 \times 10^{3}\,$m/s, an effective Bohr radius of $a_P=1.22\,$nm, and inter-donor distance
$d=10.32\,$nm. The temperature is $T=50$ mK.
\vspace*{20mm}
\label{fig:specdens_si}}
\ec
\efig

\bfig
\bc
\epsfig{file=fig3.eps, width=8.5cm, keepaspectratio=true}
\caption{(Color online) Check of convergence with respect to the memory time $\tau_{mem}=K\Delta t$ 
for the decoherence rate $\gamma$ (symmetric qubit $\epsilon=0$) and for 
the donor distances
$d=4.34\,$nm, $d=7.06\,$nm and $d=10.32\,$nm and the corresponding tunnel
matrix elements obtained from the LCAO. The Trotter time increment is fixed
to $\Delta t=0.55\omega_c$.
\label{fig:cut-off}}
\ec
\efig

\bfig
\bc
\epsfig{file=fig4.eps, width=8.5cm, keepaspectratio=true}
\caption{(Color online) Example of the Trotter convergence for the the population 
difference of the qubit, $P(t_{fix})$, from which the quantities of interest 
are extracted. In the lower sketch the tunnel amplitude was chosen as $\Delta/\omega_c=2.24$ 
and $t_{fix}=34.1\omega_c$, and
in the upper sketch $\Delta/\omega_c=3.24$ and $t_{fix}=18.2\omega_c$. The memory-time is 
fixed to $\tau_{mem}=3.85/\omega_c$ 
and three values of $K=10,11,12$ have been chosen. 
At $\tau_{mem}^2/K^{2}\rightarrow 0$ the value $P(t_{fix})$ is shown
as a result of the extrapolation $\Delta t\rightarrow 0$, with the error of the 
linear regression (horizontal bars). 
\vspace*{20mm}
\label{fig:trotter}}
\ec
\efig

\bfig
\bc
\epsfig{file=fig5.eps, width=8.5cm, keepaspectratio=true}
\caption{(Color online) Quality factor as a function of the tunneling amplitude $\Delta$ for different 
donor combinations and a small donor distance $d=4.34\,$nm. The symbols depict the exact 
QUAPI results while the dashed lines mark the results of the WCA. 
Temperature is fixed at $T=50$ mK.
\vspace*{20mm}
\label{fig:qfac1}}
\ec
\efig

\bfig
\bc
\epsfig{file=fig6.eps, width=8.5cm, keepaspectratio=true}
\caption{(Color online) Same as Fig.\ \ref{fig:qfac1}, but for an intermediate donor distance 
$d=7.06$ nm. 
\vspace*{20mm}
\label{fig:qfac2}}
\ec
\efig

\bfig
\bc
\epsfig{file=fig7.eps, width=8.5cm, keepaspectratio=true}
\caption{(Color online) Same as Fig.\ \ref{fig:qfac1}, but for a large donor 
distance $d=10.32$ nm.
\vspace*{20mm}
\label{fig:qfac3}}
\ec
\efig

\bfig
\bc
\epsfig{file=fig8.eps, width=8.5cm, keepaspectratio=true}
\caption{(Color online) Upper panel: Relaxation ($\gamma_r$) and decoherence ($\gamma$) rate for 
increasing bias $\epsilon$. Symbols are the exact QUAPI results while the dashed 
lines are the corresponding WCA results. Lower panel: Oscillation frequency $\Omega$ 
and asymptotic value $P_\infty$. The remaining parameters are 
$d=7.06$ nm, $\Delta=0.57 \omega_c$, and $T=50$ mK. 
\label{fig:bias}}
\ec
\efig

\end{document}